\documentclass[10pt]{article}
\usepackage[a4paper,margin=1.8cm]{geometry}
\usepackage{graphicx}
\usepackage{cite}
\usepackage{amsmath,amssymb,bm}
\usepackage[T1]{fontenc}
\usepackage[utf8]{inputenc}
\usepackage{microtype}
\usepackage{placeins}
\usepackage{flafter}
\usepackage[hidelinks]{hyperref}
\newcommand{\maybegraphic}[2][]{%
  \IfFileExists{#2}{\includegraphics[#1]{#2}}{%
    \fbox{\parbox[c][35mm][c]{0.9\linewidth}{\centering Missing figure file\\\texttt{#2}}}%
  }%
}

\newcommand{\etal}{et al.}

\begin{document}

\title{Reduced-order non-self-consistent Monte Carlo simulation of a planar magnetron discharge: electron heating, recapture and racetrack formation}

\author{%
F.~F. Locker$^{1}$, G. Strau{\ss}$^{2}$\\[0.6em]
\parbox{0.96\textwidth}{\centering\small
$^{1}$ Department for Ion Physics and Applied Physics, University of Innsbruck,\\ Technikerstrasse 25, 6020 Innsbruck, Austria\\
$^{2}$ Arbeitsbereich f\"ur Materialtechnologie, Institut f\"ur Konstruktion und Materialwissenschaften,\\ Universit\"at Innsbruck, Technikerstrasse 13, 6020 Innsbruck, Austria}\\[0.3em]
\normalsize\texttt{franz.locker@uibk.ac.at}}

\date{}

\twocolumn[
\begin{@twocolumnfalse}
\maketitle

\begin{abstract}
A reduced-order non-self-consistent Monte Carlo model is presented for a circular planar magnetron discharge in argon. The model combines two magnetic-field representations---a superposition of magnetic dipoles and a numerically integrated field of the finite permanent magnets---with a prescribed one-dimensional sheath--bulk potential, adaptive fourth-order Runge--Kutta orbit integration, and a null-collision treatment of electron--argon collisions. The collision module reproduces the dependence of the electron drift velocity on reduced electric field, but overestimates its absolute value by approximately a factor of 1.5; the resulting transport predictions are therefore interpreted semi-quantitatively. Applied to a magnetron geometry based on published Langmuir-probe measurements, the simulations reproduce the qualitative emergence of a cold electron population away from the cathode while retaining a hotter near-cathode component. Electrons returning to the cathode are reflected with a prescribed probability $RC$, which controls their availability for further ionising collisions. For racetrack calculations initiated with at least $2\times10^{4}$ cathode-emitted electrons and $RC=0.5$, the finite-magnet field produces a more sharply localised erosion profile whose full width at half maximum is close to a geometric racetrack-width estimate; the dipole approximation yields a broader profile. The model is not a replacement for self-consistent PIC--MCC simulations, but a computationally light tool for comparing magnetic-field representations and analysing electron heating, ionisation localisation and racetrack formation.
\end{abstract}
\vspace{1.0em}
\end{@twocolumnfalse}
]

\section{Introduction}
Magnetron sputtering is an established physical-vapour-deposition technique that combines comparatively high deposition rates, moderate substrate heating and broad materials flexibility, and can be scaled from laboratory sources to industrial coating systems \cite{HISTORY,MILESTONES,FreyKhan,THORNTON,Gudmundsson2020}. In a planar dc magnetron, crossed electric and magnetic fields confine energetic electrons close to the cathode and thereby increase the probability of ionising the working gas \cite{FreyKhan,PlanarMag,ROSSNAGEL}. The same confinement localises ion production and ion bombardment, producing the characteristic erosion racetrack. Racetrack localisation limits target utilisation and changes the spatial distribution of sputtered material, so that magnetic design is coupled directly to source lifetime, deposition rate and coating uniformity \cite{ROTMAG,RAGGL,Zhu2023Uniformity}.

Magnetron sputtering is consequently modelled with a hierarchy of methods rather than with a single universal simulation strategy. At the high-fidelity end, particle-in-cell simulations with Monte Carlo collisions (PIC--MCC) evolve charged particles together with a self-consistent electrostatic field. They can resolve non-Maxwellian electron populations, sheath formation, cross-field transport and the sensitivity of the discharge to magnetic topology \cite{Kondo1999,Kolev2006PICMCC,Kolev2006Detailed,Bogaerts2009Review,Jo2023PIC}. Such calculations provide the most direct kinetic description considered here, but their spatial and temporal resolution requirements make realistic three-dimensional geometries, broad parameter scans and process-lifetime calculations expensive.

Reduced, hybrid and multiscale approaches trade part of this self-consistency for speed and interpretability. The process is commonly separated into plasma generation, ion bombardment and sputter emission at the target, gas-phase transport of sputtered particles, and growth at the substrate \cite{Kadlec2007,Jimenez2014,Bogaerts2009Review,Brault2023Review}. Fluid or hybrid kinetic--fluid models are used for bulk plasma properties, test-particle and test-electron Monte Carlo models for energetic-particle transport, binary-collision methods for ion--solid cascades, Monte Carlo or DSMC-like methods for neutral transport, and molecular dynamics or kinetic Monte Carlo for atomistic film growth \cite{Kadlec2007,TRIM,Brault2023Review,Zhu2023Uniformity}. These modular calculations already address deposition rates, thickness distributions and film morphology, but the input fluxes and boundary conditions are often supplied by a separate model or by experiment.

Target erosion lies between the discharge and deposition stages. Self-consistent kinetic calculations have predicted ion-flux and erosion localisation for fixed target geometries, while reduced process models have related erosion to magnetic-field distributions, sputter yields, target poisoning and redeposition \cite{Kolev2006Detailed,Jo2023PIC,Strijckmans2015Erosion,Depla2019ReactiveChallenges}. A continuously coupled target-lifetime calculation would additionally update the target topography, magnetic and electric fields, plasma state, sputtered-particle transport and deposited film as erosion proceeds. Such closed-loop calculations are not yet routine, and process studies therefore still benefit from models that answer more limited ``what-if'' questions at substantially lower cost.

The near-cathode sheath is a central difficulty in constructing such a reduced model. Depending on collisionality and magnetic-field orientation, the wall transition contains a quasineutral presheath, a magnetic presheath (or Chodura sheath) and a Debye sheath, with ion--neutral collisions modifying their extent and potential drops \cite{BOHM,CHODURA,MORITZ}. In a magnetron, both the magnitude of the field and its angle to the target vary radially. Fluid and kinetic studies consequently find that the sheath width, electron confinement and potential structure vary across the cathode rather than following one universal one-dimensional profile \cite{Bradley1997,Bultinck2008,SHERIDAN2}. A prescribed effective sheath--bulk potential is therefore a strong approximation, but it is also a practical way of isolating the influence of electron transport and magnetic geometry.

Reduced electron Monte Carlo models remain useful in this regime because they retain orbit dynamics and energy-dependent collisions while avoiding a field solve at every time step \cite{Kadlec2007,Bogaerts2009Review,Jimenez2014}. Their purpose is not an absolute prediction of plasma density or target lifetime, but rapid comparison of magnetic configurations and identification of mechanisms controlling electron heating, ionisation localisation and erosion.

The present work follows this reduced-order strategy. It examines how a dipole approximation and a finite-magnet field calculation change electron confinement and the predicted racetrack, and how cathode return and reflection alter the generation-to-generation multiplication of ionising electrons. The scope is deliberately narrower than a self-consistent digital twin: the results are interpreted as qualitative and semi-quantitative process trends rather than as a complete target-evolution prediction.

\section{Model and numerical method}
\subsection{Model assumptions and scope}
The simulation describes a stationary argon discharge in a circular planar magnetron. The electrostatic potential is prescribed and time independent, its radial variation is neglected, and ignition dynamics are not modelled. Coulomb collisions between charged particles are omitted. Neutral argon is stationary, spatially uniform and represented through energy-dependent electron-collision cross-sections; effective excitation channels are used instead of a complete state-resolved excitation set.

Ions are treated as unmagnetised, and ion--neutral collisions are neglected in the erosion stage. The erosion result is therefore a reduced mapping from the simulated ion-production and cathode-impact statistics to an energy-dependent sputter yield, not a detailed kinetic prediction of ion transport or collision cascades inside the solid. These assumptions keep the calculation feasible on workstation hardware while preserving the link between magnetic confinement, electron-impact ionisation and erosion localisation.

For clarity, the initial electrons launched from the cathode are termed \emph{seed electrons}; electrons created in gas-phase ionisation events are termed \emph{ionisation-born electrons}; and electrons emitted from the cathode following ion impact are termed \emph{ion-induced secondary electrons}. The model is suitable for trend studies and comparison of magnetic configurations, but it is not used to claim self-consistent sheath structure, absolute plasma density or long-time target evolution.

\subsection{Magnetic and electric field description}
The magnetic field is calculated in two ways. In the simplified representation, the magnetron is approximated by a superposition of magnetic dipoles. In the geometrically resolved representation, the field is obtained by numerical integration over the finite magnet volumes via the magnetic vector potential,
\begin{equation}
\mathbf{A}(\mathbf{r})=\frac{\mu_0}{4\pi}\int_V \frac{\mathbf{M}(\mathbf{r}')\times(\mathbf{r}-\mathbf{r}')}{|\mathbf{r}-\mathbf{r}'|^3}\,dV',
\label{eq:Afield}
\end{equation}
in SI units, with magnetisation $\mathbf{M}=d\mathbf{m}/dV$ and $\mathbf{B}=\nabla\times\mathbf{A}$. The finite-magnet calculation is more expensive, but avoids some local artefacts of the point-dipole approximation.

The electric field is prescribed rather than obtained from Poisson's equation. Several analytic potential profiles were tested against the qualitative form of published self-consistent and analytic magnetron-sheath solutions \cite{Kondo1999,Kolev2006Detailed,SHERIDAN2,Bultinck2008}. The calculations reported here use a one-dimensional profile consisting of a near-cathode Gaussian contribution joined to a hyperbolic-cosine bulk potential. The axial field is
\begin{equation}
E_z(z)=-\frac{d\phi_{\mathrm{tot}}}{dz},
\end{equation}
with $\mathbf{E}=(0,0,E_z)$. This captures the dominant acceleration normal to the cathode. Radial electric-field structure and feedback of the particle population on the sheath are omitted.

\subsection{Particle tracing and error control}
Electron trajectories are obtained from the Lorentz-force equation
\begin{equation}
 m_{\mathrm e}\frac{d\mathbf{v}}{dt}=q\left(\mathbf{E}+\mathbf{v}\times\mathbf{B}\right),
\label{eq:lorentz}
\end{equation}
which is rewritten as a first-order system and integrated numerically. The production calculations use an adaptive fourth-order Runge--Kutta method. Fixed-step RK4 and RK5 integrations were first compared in homogeneous magnetic fields. The adaptive RK4 scheme was retained because it limited the numerical energy drift over the time scales relevant to trapped-electron trajectories.

The code was implemented in \textit{Wolfram Mathematica} and parallelised over six CPU kernels for both particle tracing and magnetic-field integration. The implementation prioritises transparent prototyping and post-processing rather than maximum throughput, and is intended for workstation-scale parameter comparisons.

\subsection{Monte Carlo collision treatment}
Electron--neutral collisions are handled with a null-collision Monte Carlo algorithm following Vahedi and Surendra \cite{VAHEDI}. The probability of a collision during a Monte Carlo interval $\Delta t$ is
\begin{equation}
P_i = 1-\exp[-\Delta t\, v_i\sigma_T(E_i)n_t(\mathbf{x}_i)],
\label{eq:Pi}
\end{equation}
where $\sigma_T=\sigma_{\mathrm{elas}}+\sigma_{\mathrm{exc}}+\sigma_{\mathrm{ion}}$ is the total energy-dependent cross-section and $n_t$ is the neutral density. Elastic, ionisation and effective excitation cross-sections are taken from the Biagi data set distributed through LXCat \cite{BIAGI}.

A constant checking frequency is defined from an upper bound on the collision frequency; events that do not correspond to a physical channel are treated as null collisions. Elastic-scattering angles are sampled from an analytic approximation to the differential cross-section. For ionisation, the post-collision energy is divided between the outgoing electrons using the distribution proposed by Opal \etal\ \cite{OPAL}. The method is simple and directly compatible with particle-based workflows, although following the low-energy population becomes expensive when a constant collision-checking interval is used.

\subsection{Electron generations, cathode return and erosion estimate}
A generation cycle begins with seed electrons launched from the cathode. Their initial kinetic energy is assigned in the range $4.36$--$4.95$~eV. This interval was chosen from the molybdenum work-function range in the original implementation and is a numerical emission assumption rather than a microscopic secondary-electron spectrum. The electrons are then accelerated by the prescribed cathode field. After entering the ionising energy range, they are followed until their energy subsequently falls below the argon ionisation threshold, $E_{\mathrm{ion}}^{\mathrm{thresh}}=15.759611$~eV, or until another termination boundary is reached. Each ionising collision creates an ionisation-born electron that is inserted into the subsequent particle ensemble. The associated ion contributes, within the reduced ion treatment, to cathode-impact and sputtering statistics; ion impacts can in turn seed a following generation of ion-induced secondary electrons.

Electrons that return to the cathode are handled through a reflection probability $RC$. The crossing time is determined from the interpolated trajectory, and a uniformly distributed random number $u\in[0,1)$ is drawn. For $u<RC$, the electron is reflected by reversing its normal velocity component; otherwise it is absorbed, or recaptured, by the cathode. Thus, increasing $RC$ increases the fraction of returning electrons that remain available for subsequent ionising collisions. The procedure represents generation-to-generation multiplication and should not be interpreted as a time-resolved self-consistent discharge avalanche.

The radial erosion profile is estimated by weighting cathode ion impacts with an energy-dependent sputter yield based on the empirical Yamamura form \cite{Yamamura}. The simulated width is compared with the geometric estimate
\begin{equation}
\omega_{\mathrm{RT}}\approx 2\sqrt{2r_{\mathrm L}^{\mathrm e}R_c},
\label{eq:wrt}
\end{equation}
where $r_{\mathrm L}^{\mathrm e}$ is the electron Larmor radius and $R_c$ is the curvature radius of the magnetic field lines at the cathode surface.

\section{Results and discussion}
\subsection{Verification strategy}
Verification is organised hierarchically because the model is deliberately reduced and non-self-consistent. First, the collision operator is subjected to a drift-velocity sanity check with a small ensemble in homogeneous argon. Second, the bookkeeping of seed electrons, ionisation-born electrons, cathode return and ion-induced secondary generations is checked internally. Third, the magnetron results are compared with published measurements and expected physical trends. The comparison concerns the emergence of a cold electron population away from the target, the localisation of ionisation and the shape of the erosion racetrack.

These checks do not constitute validation of a self-consistent plasma model. They establish the numerical logic of the reduced ingredients and define the level at which comparisons between magnetic-field representations and cathode-return assumptions can be interpreted.

\subsection{Verification against electron drift in inert argon}
The collision module was first tested in a homogeneous drift configuration. Electrons were launched in argon and accelerated by a uniform field $E_z=500$~V~m$^{-1}$. The established drift velocity after $\tau=2.93\times10^{-5}$~s was evaluated from an ensemble of six independent electron histories over $E/n_{\mathrm{Ar}}=34$--$600$~Td and compared with the published data compiled by Majeed \etal\ \cite{MAJEEB}.

The calculation reproduces the variation of drift velocity with reduced electric field, but the absolute values are systematically high by approximately a factor of 1.5. The most plausible source within the present model is the use of effective excitation channels, which underestimates the state-resolved energy loss. Because the ensemble is also small, this test is interpreted as a numerical and order-of-magnitude sanity check rather than a precision transport benchmark. The discrepancy motivates the semi-quantitative interpretation adopted throughout the paper.

\begin{figure}[!htbp]
\centering
\maybegraphic[width=0.98\linewidth]{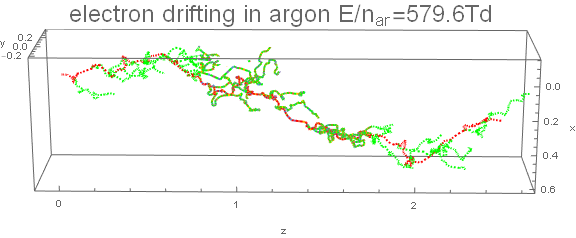}\par\medskip
\maybegraphic[width=0.98\linewidth]{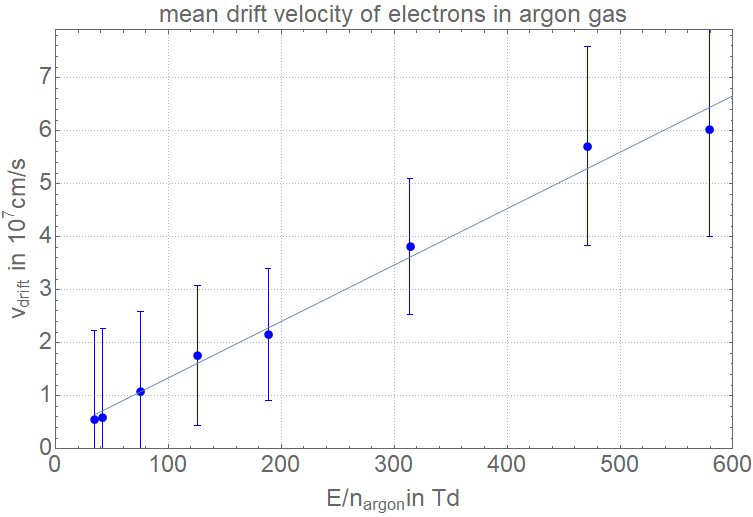}
\caption{Inert-gas drift test. Upper panel: representative electron trajectory with ionisation-born electrons in a homogeneous electric field. Lower panel: simulated mean drift velocity as a function of reduced electric field.}
\label{fig:driftvalidation}
\end{figure}

\begin{figure}[!htbp]
\centering
\maybegraphic[width=0.95\linewidth]{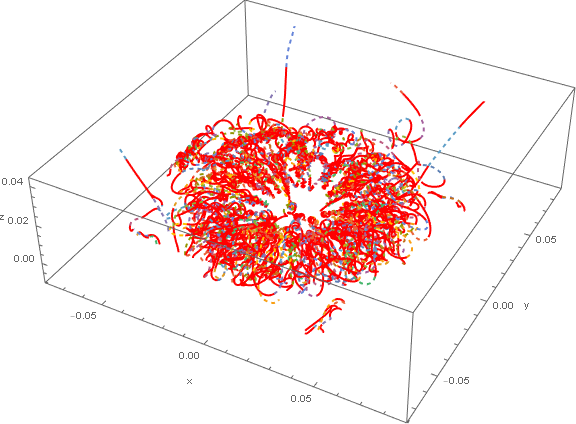}
\caption{Internal bookkeeping diagnostic illustrating the generation-cycle logic, in which seed electrons, ionisation-born electrons and ion-induced secondary electrons are followed in sequence.}
\label{fig:driftandcycle}
\end{figure}

\subsection{Electron multiplication and cathode-return diagnostics}
The internal event bookkeeping was checked by following the particle generations through ionising collisions and cathode impacts. The branching is avalanche-like only in a generation-counting sense: an ionisation-born electron may create further ionisation, and an ion impact may seed the next cathode-emitted generation. The sequence terminates when no tracked electron retains sufficient energy to create another ionisation event.

\begin{figure}[!htbp]
\centering
\maybegraphic[width=0.98\linewidth]{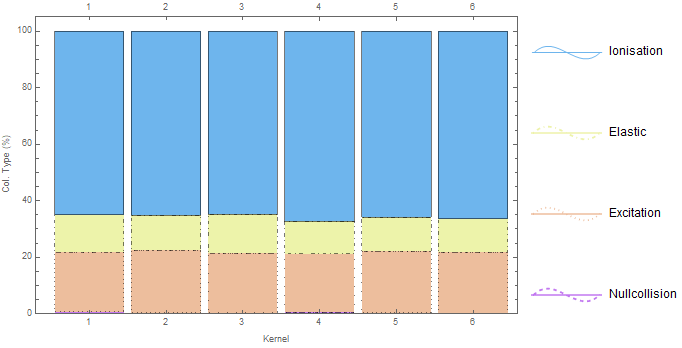}\par\medskip
\maybegraphic[width=0.98\linewidth]{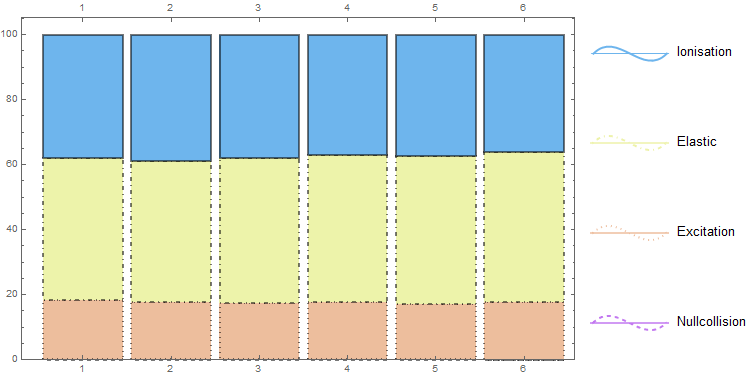}
\caption{Numbers of reflected and absorbed cathode-return events. Upper panel: dipole approximation. Lower panel: finite-magnet field.}
\label{fig:recapbars}
\end{figure}

\begin{figure}[!htbp]
\centering
\maybegraphic[width=0.98\linewidth]{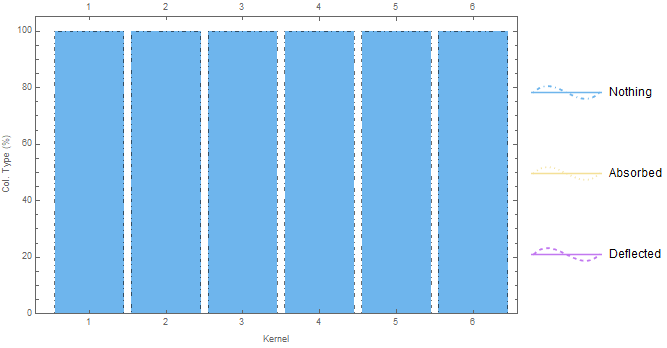}\par\medskip
\maybegraphic[width=0.98\linewidth]{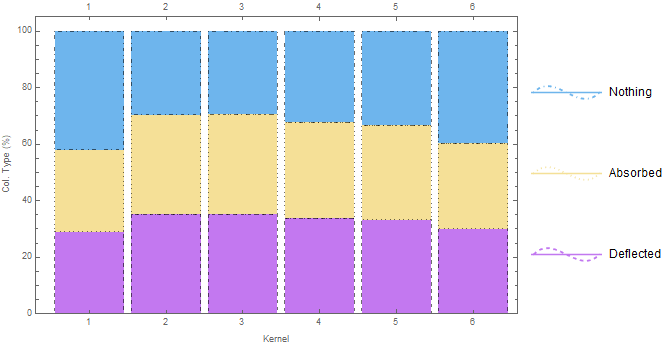}
\caption{Spatial distributions of cathode-return events in the dipole-field calculation. Upper panel: seed electrons. Lower panel: ionisation-born or ion-induced secondary electrons.}
\label{fig:recapdiagnostics}
\end{figure}

\begin{figure}[!htbp]
\centering
\maybegraphic[width=0.95\linewidth]{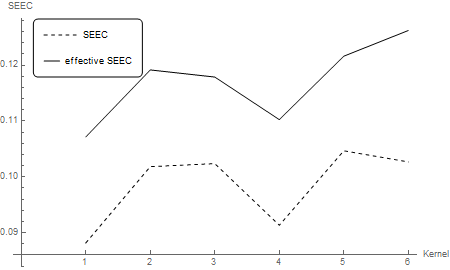}
\caption{Generation-counting diagnostic for effective secondary-electron production and cathode return in the dipole-field calculation.}
\label{fig:seecdiag}
\end{figure}

The diagnostics in figures~\ref{fig:driftandcycle}--\ref{fig:seecdiag} are internal consistency checks rather than experimental validation. They confirm that the code separates the particle classes and applies the cathode boundary rule consistently. They also show that the finite-magnet and dipole fields lead to different numbers and locations of cathode-return events. This matters for the erosion calculation because magnetic topology and cathode reflection jointly determine how many energetic electrons remain available to ionise the gas.

\subsection{Electron temperature structure}
The magnetised discharge was evaluated in a circular planar geometry based on the configuration used for the published Langmuir-probe measurements of Raggl \etal\ \cite{RAGGL}. The central magnet diameter was 20~mm and the inner and outer diameters of the ring magnet were 21 and 31~mm, respectively. No measured field map was available for the experimental source, so the magnetisation was inferred from the nominal remanence, $B_r\approx1.37$~T. This uncertainty prevents a strict point-by-point comparison of the calculated and experimental radial profiles.

For the temperature calculation, electrons were followed down to 0.1~eV and removed when they left the domain or reached a virtual measurement plane at $z=40$~mm. The ensemble contained 1200 seed electrons and approximately 1400 ionisation-born electrons. Temperature statistics were formed by radial binning and by separating axial regions below and above $z=10$~mm.

The calculation produces a hot near-cathode component and a colder population at larger axial distance. A threshold of 0.5~eV was used only as a diagnostic separation between the two plotted populations. The resulting trend is consistent with the two-temperature interpretation reported by Rossnagel and Kaufman and by Sheridan \etal\ \cite{ROSSNAGEL,SHERIDAN}, and with the increase of the cold component in the measurements of Raggl \etal\ \cite{RAGGL}. The agreement is qualitative: the hot component and axial trend are reproduced more convincingly than the detailed radial positions of the maxima. The limitations are consistent with the imposed potential, modest particle statistics and uncertainty in the experimental magnetic field.

Within this model, the cold component appears after electrons leave the high-gradient near-cathode region and enter the nearly flat bulk potential. The result therefore supports the use of the prescribed sheath--bulk profile for qualitative electron-heating studies, but does not validate it as a self-consistent sheath solution.

\begin{figure}[!htbp]
\centering
\maybegraphic[width=0.98\linewidth]{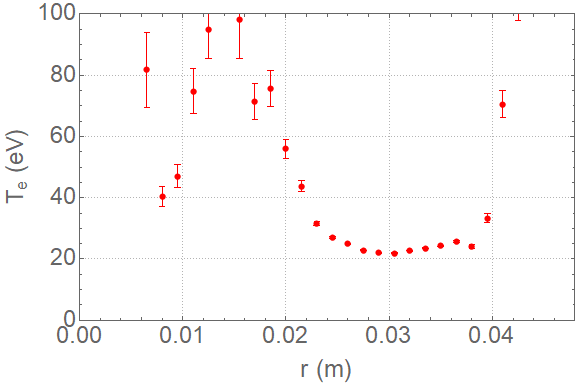}\par\medskip
\maybegraphic[width=0.98\linewidth]{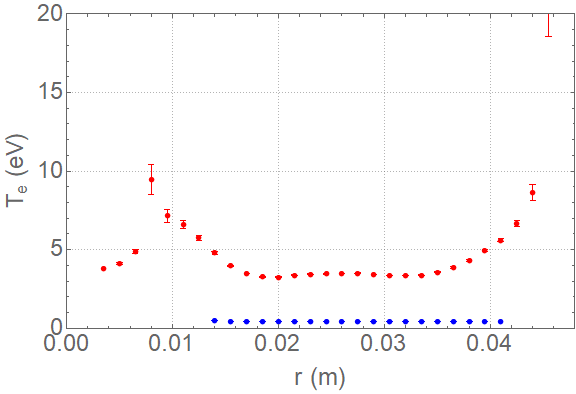}
\caption{Simulated radial mean electron-temperature distributions. Upper panel: close to the cathode. Lower panel: near the virtual measurement plane.}
\label{fig:etempsim}
\end{figure}

\begin{figure}[!htbp]
\centering
\maybegraphic[width=0.95\linewidth]{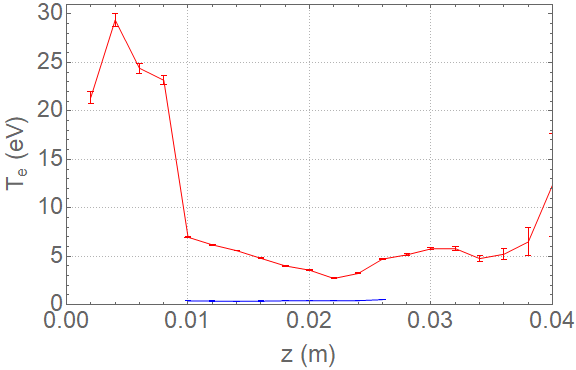}
\caption{Simulated axial trend of the electron temperature. Published measurements by Raggl \etal\ \cite{RAGGL} and earlier literature trends \cite{ROSSNAGEL,SHERIDAN} are used for comparison in the text but are not reproduced as separate figures.}
\label{fig:etempz}
\end{figure}

\subsection{Magnetic-field representation and ionisation localisation}
The dipole approximation and the finite-magnet integration produce visibly different confinement regions. The point-dipole representation is inexpensive and suitable for exploratory scans, but it distorts the local field topology near the cathode centre and broadens the region in which energetic electrons remain confined. The finite-magnet field yields a more compact ionisation distribution and a narrower radial concentration of cathode impacts.

Because no measured field map is available, this comparison does not establish absolute field accuracy. It does show that geometric resolution of the permanent magnets is a first-order modelling choice for the predicted ionisation and racetrack widths; increasing complexity elsewhere cannot compensate for a qualitatively distorted magnetic topology.

\begin{figure}[!htbp]
\centering
\maybegraphic[width=0.98\linewidth]{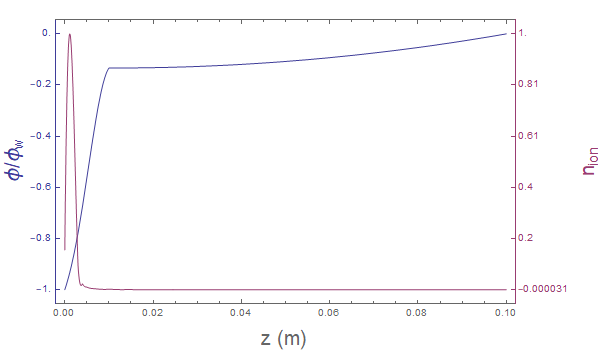}\par\medskip
\maybegraphic[width=0.98\linewidth]{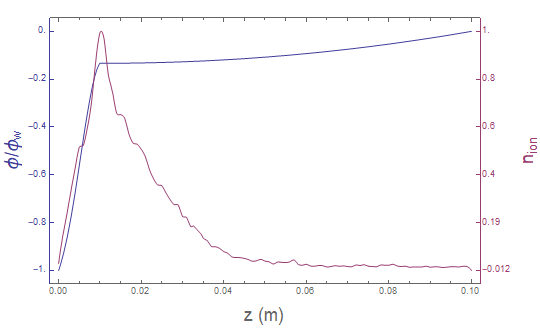}
\caption{Normalised axial dependence of ionisation events and the prescribed potential. Upper panel: finite-magnet field. Lower panel: dipole approximation.}
\label{fig:ionz}
\end{figure}

\subsection{Cathode return and reflection}
The return boundary condition distinguishes electrons absorbed by the cathode from those reflected back into the discharge. With the present definition, $RC$ is the reflection probability, while $1-RC$ is the probability of absorption or recapture. Increasing $RC$ therefore increases the population available for renewed ionising collisions and strengthens the generation-to-generation multiplication represented by the model.

This parameter is not presented as a universal material coefficient. It combines surface reflection and unresolved cathode-interaction physics into one controllable boundary condition. Its influence is nevertheless physically relevant: magnetic confinement determines how often electrons return to the target, while $RC$ determines what fraction of those returns terminates a trajectory. Similar sensitivity to cathode recapture has been found in self-consistent planar-magnetron calculations \cite{RECAP}.

\subsection{Racetrack prediction}
For each magnetic-field representation, the erosion calculation was initiated with at least $N_e\ge2\times10^{4}$ seed electrons and used $RC=0.5$. The geometry was chosen to reproduce the published Raggl configuration as closely as possible, subject to the uncertainty in the actual magnetisation and field map.

Cathode ion impacts were weighted by the energy-dependent sputter yield and accumulated radially. For the finite-magnet field, the profile rises sharply from its baseline to half maximum and the predicted full width at half maximum is close to the geometric estimate $\omega_{\mathrm{RT}}$ in equation~(\ref{eq:wrt}). The dipole calculation produces a broader transition and erosion band. The published IFM surface image of the eroded molybdenum target reported by Raggl \etal\ \cite{RAGGL}, which is not reproduced here, also shows a sharply localised trench. This comparison is qualitative because the measured depth profile and magnetic-field map were not available as numerical inputs.

The result should therefore be read as a discrimination test between magnetic-field representations, not as a self-consistent target-lifetime simulation. The surface shape is not updated and the discharge is not recomputed as the target erodes. Within that reduced scope, however, the calculation shows that resolving the finite magnet geometry is necessary to obtain the observed degree of racetrack localisation.

\begin{figure}[!htbp]
\centering
\maybegraphic[width=0.98\linewidth]{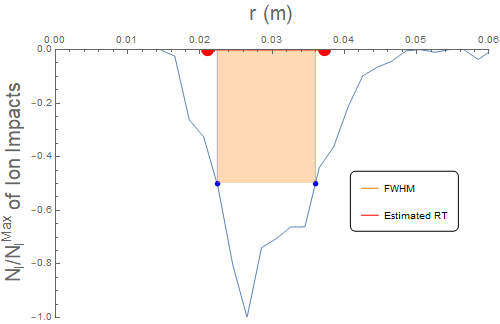}\par\medskip
\maybegraphic[width=0.98\linewidth]{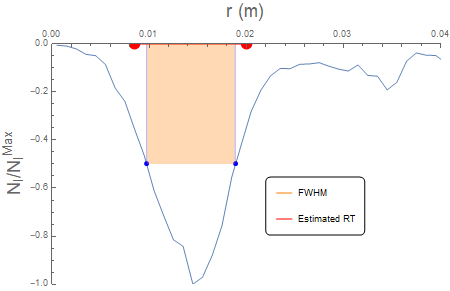}
\caption{Simulated radial erosion profiles including the geometric racetrack-width estimate $\omega_{\mathrm{RT}}$. Upper panel: finite-magnet field. Lower panel: dipole approximation.}
\label{fig:rtsim}
\end{figure}

\FloatBarrier
\section{Conclusions}
A reduced-order, non-self-consistent electron Monte Carlo model was used to compare two magnetic-field representations in a circular planar magnetron. The prescribed sheath--bulk potential and collision model reproduce qualitative features of electron heating, including a hot near-cathode population and a colder component farther from the target. The drift test, however, overestimates the absolute electron drift velocity by about a factor of 1.5, so the model should be regarded as semi-quantitative.

The clearest process-facing result is the sensitivity of the predicted racetrack to the magnetic-field representation. The point-dipole approximation broadens the confinement and erosion regions. Numerical integration over the finite magnets produces more localised ionisation and a sharper erosion band with a width close to the geometric racetrack estimate and qualitatively consistent with the published experimental trench. This identifies magnetic-field geometry as the dominant modelling requirement for the present reduced erosion calculation.

Cathode return is also consequential. The reflection probability $RC$ controls whether returning electrons are lost or remain available for additional ionising collisions, and therefore changes the generation-to-generation multiplication represented by the model. Because $RC$ combines unresolved surface processes, conclusions based on its value should be interpreted as sensitivity results rather than material predictions.

The model does not solve Poisson's equation, include collisional ion transport, or update the target surface as erosion proceeds. It therefore cannot predict absolute plasma density or target lifetime. Its utility is narrower: inexpensive comparison of magnetic configurations and identification of mechanisms that localise ionisation and erosion. Quantitative extension would require measured magnetic-field and erosion profiles, particle-number convergence studies, a constrained cathode-interaction model, and eventual coupling to a self-consistent or hybrid electric-field description.

\section*{Author contributions}
\noindent\textbf{F.~F. Locker}: Conceptualization, methodology, software, validation, formal analysis, data curation, visualization, project administration, resources, and writing---original draft.

\noindent\textbf{G. Strau{\ss}}: Conceptualization, supervision, resources, and project administration.

\section*{Data availability statement}
The data that support the findings of this study are available from the corresponding author upon reasonable request.

\section*{Acknowledgements}
The authors thank David Stock for valuable discussion and acknowledge the use of computational resources at the University of Innsbruck. The Authors further thank Nikolaus Weinberger and Alexander Kendl for valuable discussion.

\FloatBarrier
\bibliographystyle{unsrt}
\bibliography{Bibli_arxiv_clean}

\end{document}